\documentclass[conference,10pt]{IEEEtran}
\usepackage{epsfig,rotating,setspace,latexsym,amsmath,epsf,amssymb,amsfonts,bm,theorem,subfigure,epstopdf}
\usepackage{cite,authblk}
\usepackage{bbm}
\usepackage{color}
\usepackage{mathtools}
\usepackage{algorithm}
\usepackage{algpseudocode}
\usepackage{verbatim}
\usepackage{xcolor}

\algrenewcommand\algorithmicforall{\textbf{foreach}}
\algrenewcommand\algorithmicindent{.8em}

\algnewcommand\algorithmicforeach{\textbf{for each}}
\algdef{S}[FOR]{ForEach}[1]{\algorithmicforeach\ #1\ \algorithmicdo}

\IEEEoverridecommandlockouts
\allowdisplaybreaks

\begin{document}

\title{Deep Learning-Based Real-Time Rate Control for Live Streaming on Wireless Networks}

\author{
Matin Mortaheb$^{\dag}$, Mohammad A. (Amir) Khojastepour$^{*}$, Srimat T. Chakradhar$^{*}$, Sennur Ulukus$^{\dag}$ \\
\normalsize $^{\dag}$University of Maryland, College Park, MD, $^{*}$NEC Laboratories America, Princeton, NJ \\
\normalsize \emph{mortaheb@umd.edu, amir@nec-labs.com, chak@nec-labs.com, ulukus@umd.edu}
}
\maketitle

\begin{abstract}
Providing wireless users with high-quality video content has become increasingly important. However, ensuring consistent video quality poses challenges due to variable encoded bitrate caused by dynamic video content and fluctuating channel bitrate caused by wireless fading effects. Suboptimal selection of encoder parameters can lead to video quality loss due to underutilized bandwidth or the introduction of video artifacts due to packet loss. To address this, a real-time deep learning based H.264 controller is proposed. This controller leverages instantaneous channel quality data driven from the physical layer, along with the video chunk, to dynamically estimate the optimal encoder parameters with a negligible delay in real-time. The objective is to maintain an encoded video bitrate slightly below the available channel bitrate. Experimental results, conducted on both QCIF dataset and a diverse selection of random videos from public datasets, validate the effectiveness of the approach. Remarkably, improvements of 10-20 dB in PSNR with repect to the state-of-the-art adaptive bitrate video streaming is achieved, with an average packet drop rate as low as 0.002.
\end{abstract}

\section{Introduction}

The evolution of live streaming and video transmission has fundamentally reshaped how we connect, communicate, and consume content. High-speed internet and advanced technology empower individuals and organizations to instantly share real-time experiences, from live concerts and sports events to educational seminars and personal vlogs, transcending geographical boundaries and fostering immediacy and interactivity. Consequently, there is a growing emphasis on video transmission technologies that guarantee seamless, low-latency, high-quality live streaming over wireless channels.

A compelling use case for such technology is exemplified in the transmission of closed-circuit television (CCTV) video feeds via wireless 5G systems. CCTV serves diverse purposes, from providing multiple angles for live events like concerts and sports matches to securing large public areas. Leveraging the remarkable attributes of 5G wireless networks, such as, high speed, low latency, and large bandwidth, CCTV video is now exceptionally clear and responsive. However, dynamic wireless channels introduce challenges, leading to potential issues like freezing and color distortions due to varying channel quality. Therefore, adapting video bitrate to match the available channel bitrate has become imperative.

In this work, we consider standard compliant systems where video compression is executed by a video encoder (VE), such as an H.264 encoder \cite{wiegand2003overview}. The encoded video stream is then encapsulated into container frames, which are subsequently transmitted via the physical layer. Employing standard compliant VE and communication systems offers advantages such as streamlined deployment and interoperability, alongside harnessing the efficiency of specialized hardware accelerations.

Adaptive bitrate (ABR) streaming effectively addresses these challenges by dynamically optimizing the bitrate based on network conditions, ensuring a smooth and consistent viewing experience despite data rate fluctuations. However, aligning the encoded video bitrate with the available channel bitrate is highly complex. Video bitrate fluctuates due to numerous factors, including visual complexity, color dynamics, motion, and scene variations. Simultaneously, channel bitrate changes occur due to wireless channel fading. Despite the ability to control video bitrate via VE parameter adjustments, the exact relationship between output bitrate and parameters remains elusive. This complexity is especially evident in live video streams, necessitating real-time VE parameter adjustments to optimize video quality within target encoding bitrates.

\begin{figure}[]
 \centerline{\includegraphics[width=0.9\linewidth]{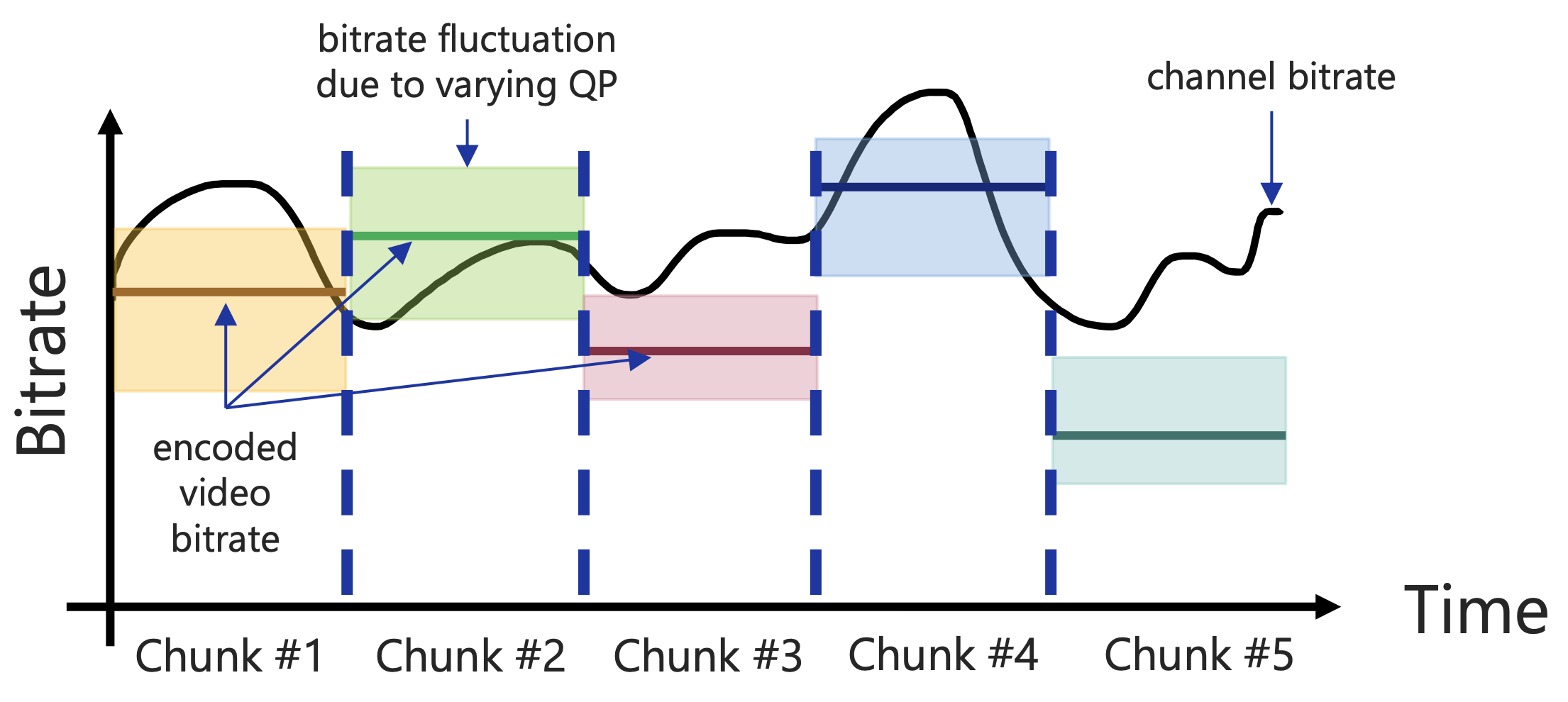}}
 \vspace*{-0.3cm}
  \caption{Fluctuations of encoded video bitrate and available channel bitrate over chunks.}
  \label{fig:fluctuation}
  \vspace*{-0.4cm}
\end{figure}

The illustration presented in Fig.~\ref{fig:fluctuation} captures the inherent variations in both video and channel bitrates. As input data is processed in chunks by the VE, the encoded video bitrate is established on a chunk-by-chunk basis. Fig.~\ref{fig:fluctuation} brings into focus the dynamic range of encoded bitrates (depicted as solid lines) across distinct chunks, showcasing this diversity despite the uniform employment of quantization parameter (QP) at $QP=20$ by the VE. Further, the encircling shaded region adjacent to the solid encoded bitrate line outlines the extent of fluctuations in the encoded video bitrate within the range of $15< QP < 25$. Lastly, Fig.~\ref{fig:fluctuation} represents the channel bitrate using a curve, reflecting the reality that channel quality index (CQI) feedback from users is often provided within a shorter time frame than that of a single chunk.

Several widely recognized adaptive bitrate streaming implementations are prevalent today, including Dynamic Adaptive Streaming over HTTP (DASH) \cite{stockhammer2011dynamic}, Apple's HTTP Live Streaming (HLS) \cite{HLS}, Microsoft's Live Smooth Streaming (Smooth) \cite{Smooth}, and Adobe's Adaptive Streaming (HDS) \cite{Adobe}. These technologies have attained significant popularity, serving as vital conduits for delivering video content over the internet. The state-of-the-art in ABR transmission, exemplified by DASH and HLS, has indubitably revolutionized content delivery over the internet. These schemes seamlessly transition between various video quality levels, culminating in uninterrupted playback experiences. However, these shifts are confined to occur between video segments, and the necessity for pre-encoded video qualities restricts their applicability in live streaming and real-time encoding scenarios. 

Compounded by the differences in video segment sizes and chunk durations, DASH's adaptability is hindered by the rapid changes in channel bitrate precipitated by wireless channel fading. Additionally, the volatile nature of video bitrates for DASH or HLS video segments complicates the adaptation of encoded bitrates to channel bandwidth variations. Consequently, the efficacy of DASH and HLS technologies in wired networks does not translate seamlessly to wireless environments, where the dynamic fluctuations of channels pose a challenge for consistent streaming experiences.

Despite their wide adoption, DASH and HLS protocols possess certain limitations that compromise their suitability for wireless channels and real-time applications, particularly live streaming scenarios. The ABR technology in DASH and HLS hinges on user-initiated and controlled selections, based on past packet receptions, device capabilities, and estimated available bitrates. This reactive approach, while delivering requested video streams, falls short in actively providing the optimal quality based on real-time network conditions. Therefore, while these technologies offer valuable adaptive bitrate streaming capabilities, their limitations warrant careful consideration when selecting streaming solutions for live streaming over wireless channels. Our research underscores that, within 5G networks, DASH and HLS incur significant penalties in video quality (about 10-20db PSNR penalty) and artifact occurrence (about 1\%-10\% of the video chunks), underpinning the need for alternative approaches.

In response, this paper introduces a network-aware real-time rate control (RTRC) framework, where an RTRC controller dynamically adjusts encoding parameters in real-time to encode individual video chunks with negligible delay. This controller accurately forecasts the encoded video bitrate, factoring in both ``input video" and $QP$ parameter. As a result, $QP$ is strategically chosen to ensure that the encoded video bitrate for each chunk remains below the anticipated user bitrate, preventing bitrate overshoot and packet drops while minimizing the likelihood of user-visible artifacts. Simultaneously, optimized $QP$ harnesses available bandwidth, thereby maximizing received video quality. 

RTRC's foundation rests on three pivotal pillars: (i) accessing user channel quality through the physical layer, (ii) forecasting the encoded video bitrates based on ``input video" and $QP$ and adapting $QP$ to maximize the encoded video quality and satisfy bitrate constraints in harmony with the available channel bitrate, and (iii) encapsulating encoded video chunks into container packets and transmitting them via the physical layer. The proposed technology finely tailors the video stream to user capabilities, gracefully accommodating network quality fluctuations and ensuring the highest attainable video quality.

\section{System Model}\label{sec:model}

We consider a transmitter that aims to send a live video feed to a receiver with the best possible resolution and minimal to no delay. The transmitter uses an standard compliant VE, signal modulator and channel encoder (MCE). We consider H.264 codec for VE and measure the video resolution in terms of average PSNR which is obtained as the mean of the PSNR of the reconstructed video frames. We assume that the video is taken and encoded with a fixed number of video frames per second ($fps$) where each video frame has constant width $W$ and height $H$. The VE has minimal delay, hence the video buffer before and after VE only contains active video frames over which the temporal encoding/decoding is performed. 

We assume that VE takes a \emph{chunk} of consecutive video frames and produces the encoded video, hence a chunk contains one or more group of pictures (GOP) over which the temporal encoding/decoding happens. The encoded video $\Tilde{V}$ for a chunk $V$ is generated by VE with particular $H$, $W$, and $fps$. The bitrate of $\Tilde{V}$, i.e., the encoded bitrate $b(V, QP)$, is a function of quantization parameter $QP$ using H.264 codec. 
The encoded video is then encapsulated into \emph{container packets} which are sent through the channel.

\begin{figure}[]
 \centerline{\includegraphics[width=0.8\linewidth]{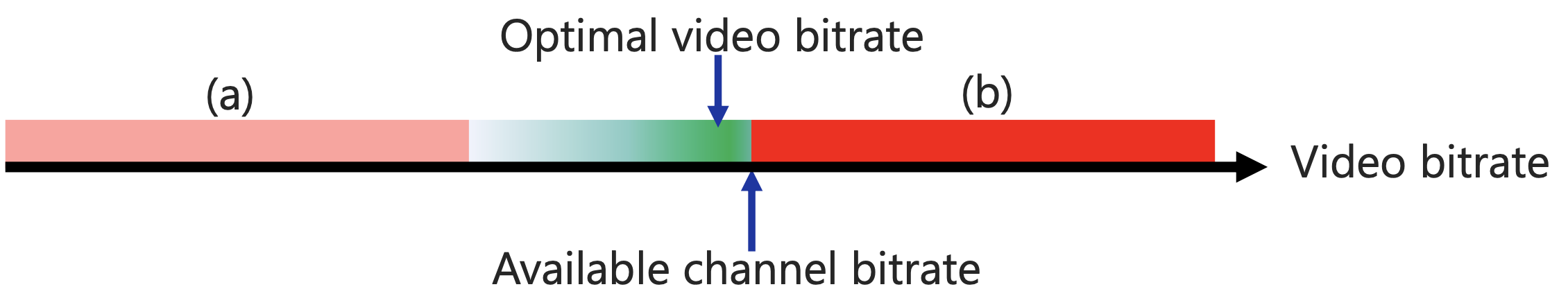}}
 \vspace*{-0.3cm}
  \caption{Consequences of choosing wrong value for QP.}
  \label{fig:wrongQO_selection}
  \vspace*{-0.4cm}
\end{figure}

We consider MCE that is compliant with modulation and coding scheme table from 3GPP/LTE standard \cite[Table~7.2.3]{3gpp_TS36_213}). The physical layer performs MCE for each container packet, ensuring error-free communication as long as the encoded rate stays below the available channel rate. We assume a Rayleigh fading channel with the channel coefficient $h$ following a circularly symmetric complex Gaussian distribution (zero mean and variance $\sigma_n^2$). The supported transmission rate for an instantaneous channel $h$ is then given by $r(h) = \log \left(1 + \frac{P |h|^2}{\sigma^2 G} \right)$, where $G$ models the gap to the capacity due to the channel coding. Please note that in this work we assume $G$ is a fixed number and do not consider its dependency on the modulation and coding scheme (MCS). Also, we do not consider discrete values for $r(h)$ due to MCS. 

We assume that the channel coherence time is longer than the duration of a chunk and the channel estimate is available for each chunk. Hence, if the encoded video rate for a chunk is not higher than the available channel rate, all the container packets will be received without error. However, if the encoded video rate exceeds the channel's supported rate, the physical layer encodes and transmits the container packets with the supported rate of the channel, hence, a number of container packets in such chunks cannot be transmitted on time by the network and they will be dropped. The number of successful packets transmitted in the duration of one chunk will be proportional to the instantaneous channel rate for that chunk.

\begin{figure}[]
 \centerline{\includegraphics[width=0.8\linewidth]{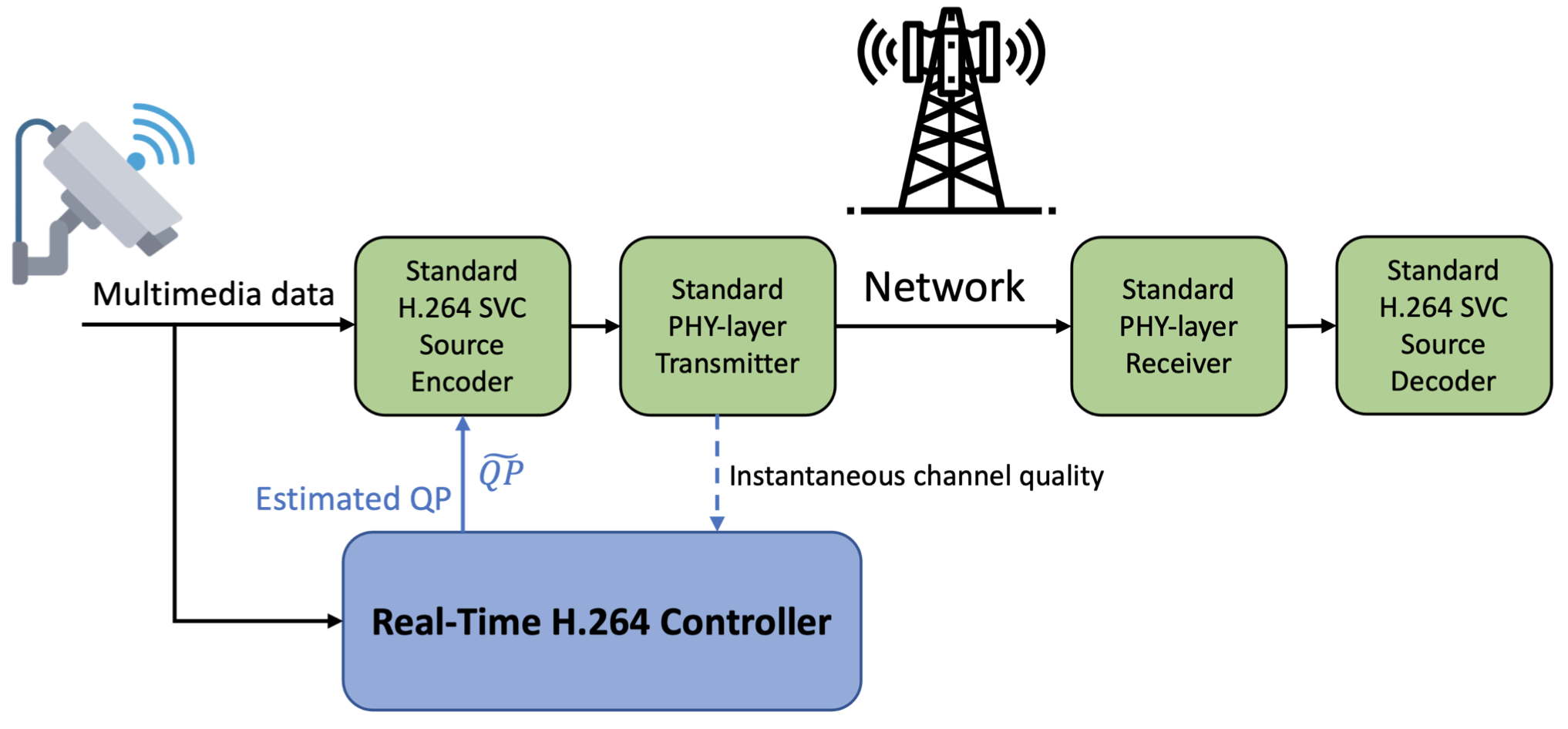}}
 \vspace*{-0.3cm}
  \caption{System model for RTRC.}
  \label{fig:system_model}
  \vspace*{-0.5cm}
\end{figure}

This variability of the channel bitrate $r(h)$ due to the fading nature of the wireless channel, i.e., $h$, as well as the variations in the video bitrates per video chunk have to be carefully considered in the design of an RTRC system. In live streaming, the RTRC should be able to identify a correct value of $QP$ that maximizes the video quality while satisfies the constraint on the encoded bitrate $b(V,QP)$ in real-time. 
If the encoder selects a $QP$ value that is too high, the resulting encoded video bitrate is too low and the constraint on the encoded video bitrate can be easily satisfied. However, a high value of $QP$ leads to poor video quality on the receiver's end and an inefficient utilization of the available bandwidth. Conversely, if the selected $QP$ is too low, the encoded video bitrate exceeds the channel bitrate, i.e., $b(V,QP) > r(h)$, that results in dropped video frames due to the channel bitrate limitations. This drop in frames leads to a noticeable decline in the received video quality due to artifacts such as freezing, blank screens, color distortions, flickering, and so on. 

Fig.~\ref{fig:wrongQO_selection} illustrates two regions for encoded video bitrates with respect to an available channel bitrate for which either the bandwidth is underutilized hence video quality is bad (denoted by (a) in Fig.~\ref{fig:wrongQO_selection}), or some packets are dropped and the video has artifacts (denoted by (b) in Fig.~\ref{fig:wrongQO_selection}). Finally, Fig.~\ref{fig:wrongQO_selection} illustrates a region for the encoded video bitrate between regions (a) and (b) where good video quality may be achieved without packet drop and artifacts. Of course, the closer the encoded bitrate to the available channel bitrate, the better the video quality.

Hence, our goal is to design a deep learning (DL) based network-aware RTRC which estimates the $QP$ value in real-time (i.e., negligible delay) that maximizes the encoded video quality for a given available channel bitrate. This $QP$ value is then used by the H.264 video encoder to generate the encoded video. The complete system model, where the proposed RTRC is highlighted with a blue bounding box, is shown in Fig.~\ref{fig:system_model}.

\section{Deep Learning Based Rate Control} \label{sec:RC}

Considering that video codecs commonly possess standardized implementations and can leverage existing hardware acceleration, the central objective of a rate control unit (RCU) resides in the precise selection of the appropriate quantization parameter, $QP$, for each video chunk that will be applied to the encoder. This meticulous calibration endeavor is directed towards attaining an optimal video quality while adhering to maximum bitrate constraint for individual video chunks. 

As discussed earlier the main challenge of the RCU is correctly forecasting the encoded video bitrates based on ``input video" and $QP$ so that the rate control (RC) module can select the $QP$ to maximize the encoded video quality and satisfy bitrate constraints in harmony with the available channel bitrate. In this work, we exploit a DL network which is proven very effective in capturing the dynamic aspects and essential information within the video, which influences the encoded bitrate of H.264 codec for different values of $QP$.

\begin{figure}[]
 \centerline{\includegraphics[width=0.8\linewidth]{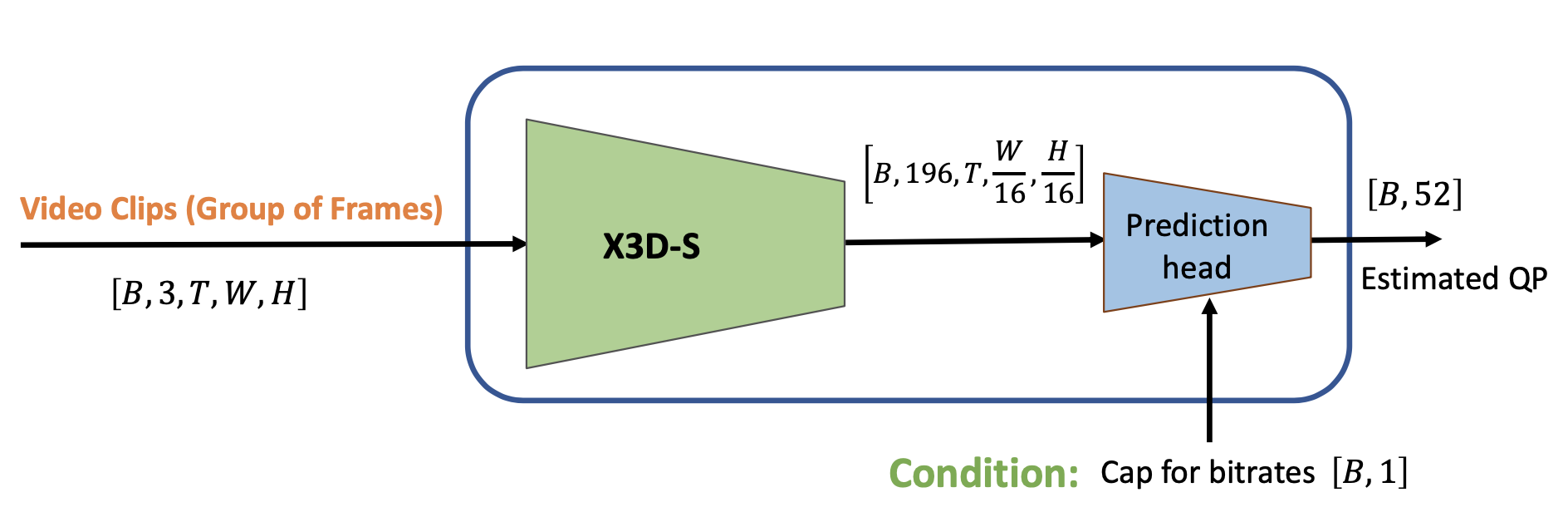}}
 \vspace*{-0.3cm}
  \caption{Model structure for the rate control unit (RCU).}
  \label{fig:RC_model}
  \vspace*{-0.5cm}
\end{figure}

Fig.~\ref{fig:RC_model} illustrates the architecture of the RCU model. The RCU receives an input video chunk along with a specified bitrate cap, denoted as $BR_{max}$. The resulting output is the $QP$ value, which falls within the range of [0, 51]. This $QP$ value is subsequently employed in the following encoding stage.

The architecture of RCU is given as follows. We employ X3D-S \cite{feichtenhofer2020x3d} as the base network for capturing video dynamics. X3D-S incorporates a multitude of convolutional layers, significantly augmenting its capability to recognize and comprehend video content. X3D-S processes video content with dimensions of $[B, 3, T, W, H]$, yielding corresponding features sized as $[B, 196, T, W/16, H/16]$, where $B$ represents the batch size, $T$ denotes the number of frames in a segment, and $W$ and $H$ stand for the width and height of the video segment.

The resulting feature is then further processed through a DNN prediction head comprising several convolutional layers, each followed by a conditional group normalization (CGN) block \cite{wu2018group}. The CGN blocks normalize the output from the previous layer. Additionally, it takes a tensor of $\log_{10}(BR_{max})$ with a size of $[B, 1]$ as a conditioning factor. Each element in this tensor represents the bitrate cap for each video within the batch. CGN processes this condition through 3 linear layers, each incorporating the Gelu activation function \cite{hendrycks2016gelu}, which transforms the tensor size to $[2B, 1]$. Subsequently, this tensor is split into two tensors of size $[B, 1]$, referred to as $\gamma$ and $\beta$. These tensors are then employed to adjust the normalized output from the preceding layer using the formula $\gamma \times output + \beta$. This approach effectively trains the video feature to discern $QP$ values across various scenarios with varying $BR_{max}$ magnitudes.

Our training-set comprises three key elements: A video chunk $V$, a specified target quantization parameter $QP_{target}$, and the corresponding bitrate $BR_{target}$ linked to the designated $QP_{target}$. In our RCU, we utilize $V$ as an input for the X3D-S model and employ the $BR_{target}$ value as a conditioning factor for the CGN blocks. The outcome generated by the RC unit is an estimated $QP$ represented as $\tilde{QP}$. We employ the following loss function,
\begin{align}
    L = L_{CE} (\Tilde{QP},QP_{target})
\end{align}
where $L_{CE}$ represents a cross-entropy loss function. 

It is worth noting that during the training of the RCU, $BR_{target}$ is not strictly treated as an upper bitrate limit. Therefore, during the testing phase, there is a possibility that the estimated $QP$ is slightly lower (usually by one or two steps) than the $QP$ value required to ensure a certain bitrate caps. To mitigate this rounding effect, we use slightly higher values for $QP$ than the estimated $QP$ value for encoding the data. This adjustment effectively prevents the occurrence of unwanted and destructive artifacts in the corresponding video chunk, while the resulting drop in video quality due to incremental change in $QP$ by one or two steps remains negligible. We note that while using a higher value of $QP$ than the estimated value by RCU improves the packet success rate and minimizes the probability of seeing an artifact, it will also result in decrease in bandwidth efficiency and lowers the overall transmitted video quality. 

\section{Performance Evaluation}\label{sec:evaluation}

The encoded video bitrate in state-of-the-art video coding standards, such as H.264 \cite{wiegand2003overview} and H.265 \cite{sullivan2012overview}, exhibits notable fluctuations. This phenomenon is well-documented, with I frames requiring significantly higher bitrates compared to P or B frames. However, these fluctuations are not limited to individual frames; they extend to larger GOPs as well.

Our extensive studies, involving hundreds of random videos from public datasets reveal that the coefficient of variation, $CV = std/mean$, computed over 10-100 consecutive  chunks of the same video has a mean of about 0.3 which remains relatively high. In this study, the duration of the video chunks is 1 second that is equivalent to 30 frames when $fps = 30$; hence, a chunk corresponds to multiple GOPs or a considerably large GOP. To visually illustrate this encoded bitrate fluctuation within a single video, we use a typical video taken from a football match and partition it into 100 one-second chunks and encode each chunk using 6 values of $QP = (1, 20, 33, 37, 44, 51)$.

\begin{figure}[]
 \centerline{\includegraphics[width=0.7\linewidth]{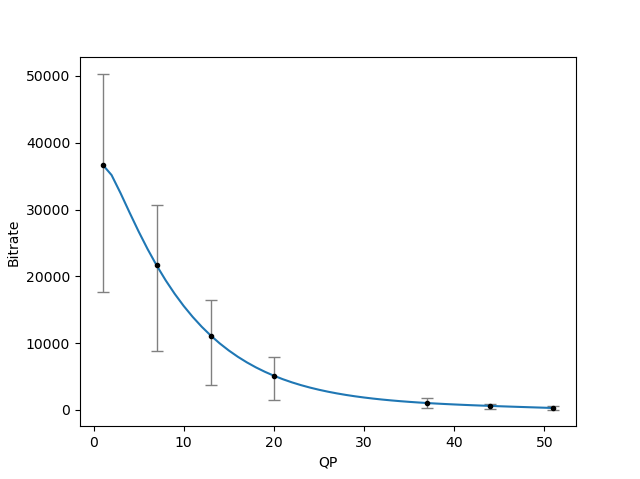}}
 \vspace*{-0.3cm}
  \caption{Variance of encoded video bitrate over chunks in different $QP$ values.}
  \label{fig:bitrate_variance}
  \vspace*{-0.3cm}
\end{figure}

Fig.~\ref{fig:bitrate_variance} illustrates the variation of the encoded video bitrate across each quality level. This observation underscores the necessity of employing deep learning techniques, as they offer an effective means to understand video dynamics within a chunk and precisely adjust parameters like $QP$ while adhering to the available channel bandwidth constraints.

\begin{figure}[]
 	\begin{center}
 	\subfigure[]{%
 	\includegraphics[width=0.6\linewidth]{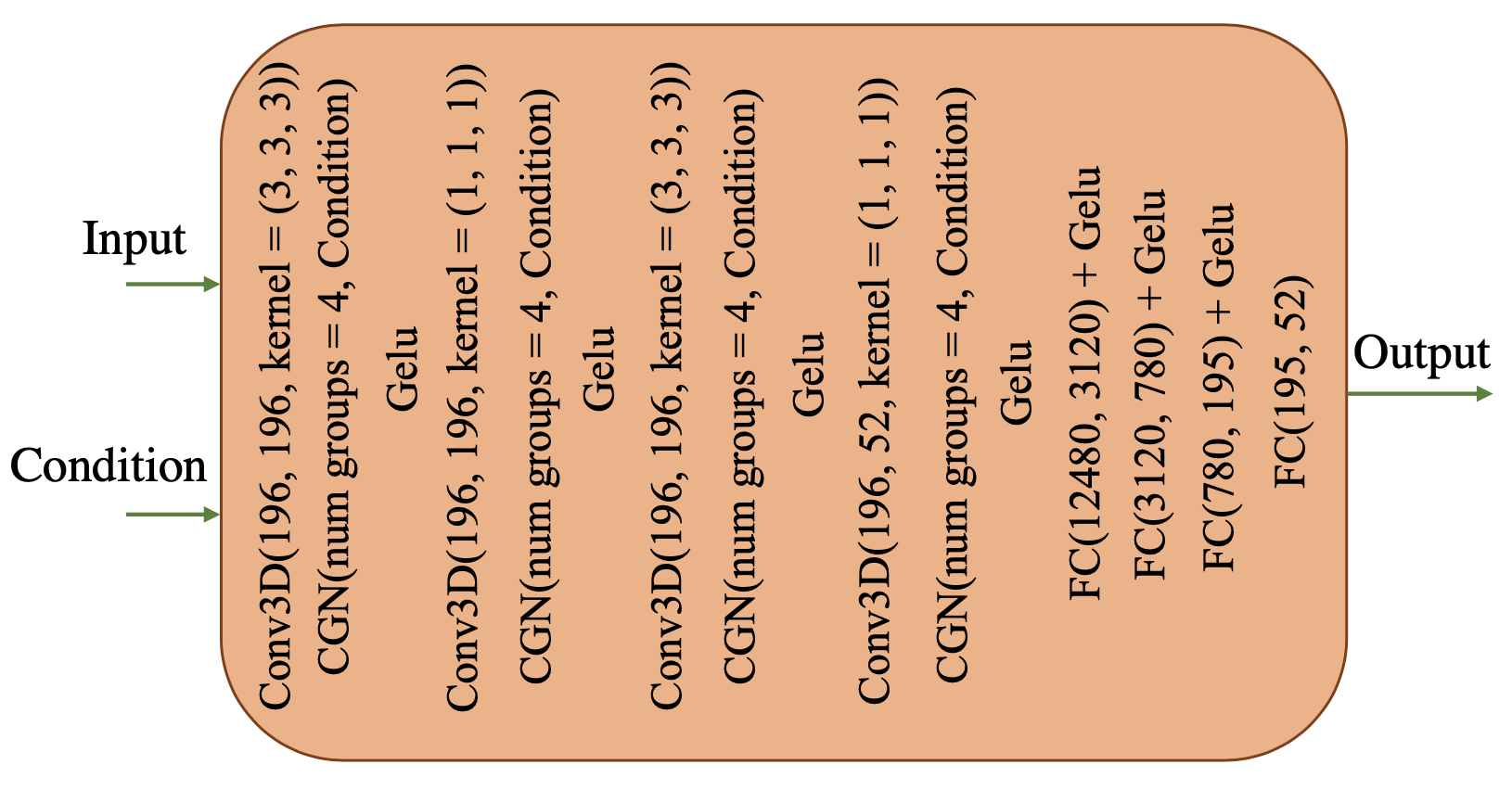}}
 	\subfigure[]{%
 	\includegraphics[width=0.38\linewidth]{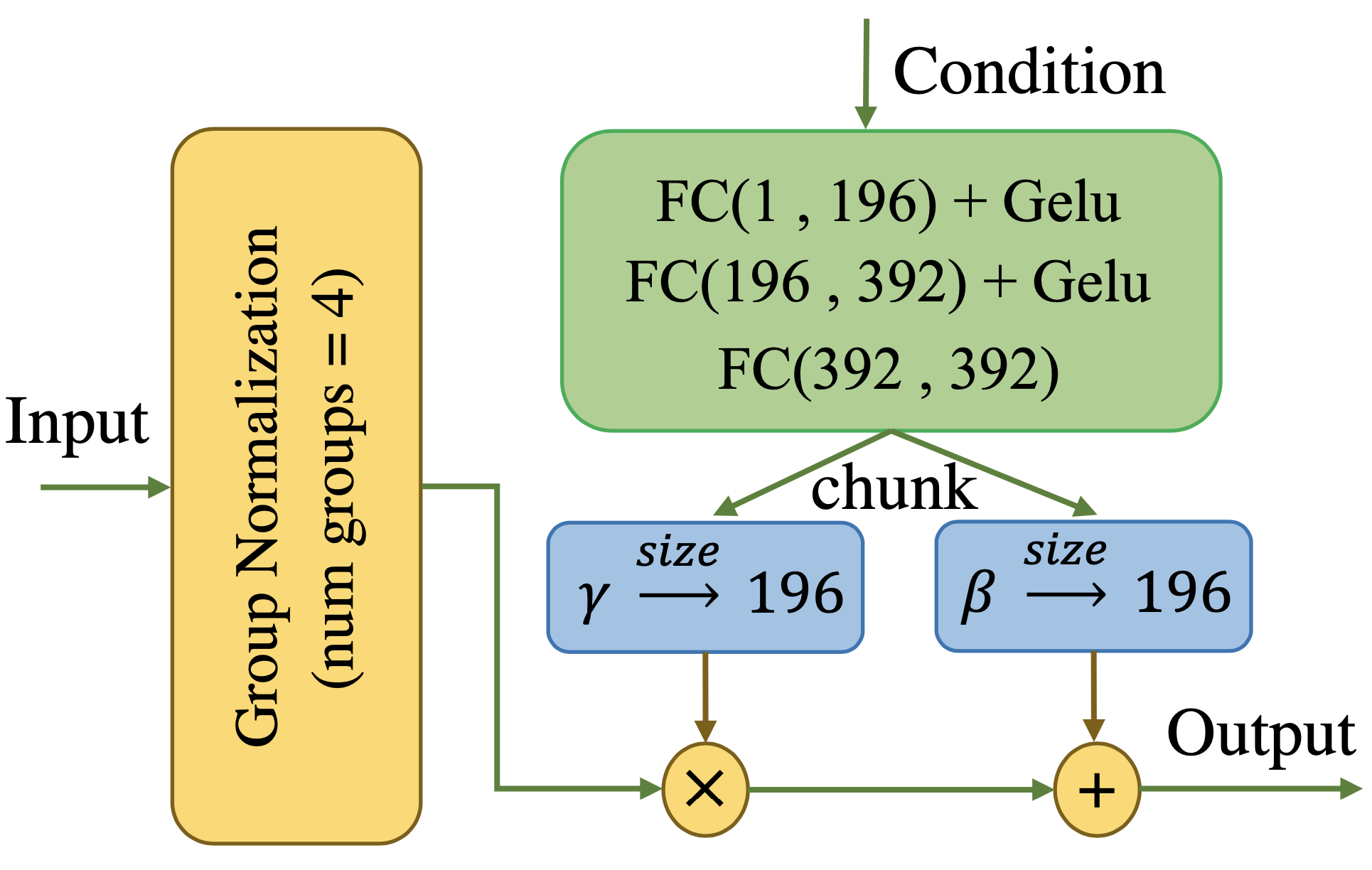}}\\ 
 	\end{center}
    \vspace*{-0.4cm}
 	\caption{Model structure for (a) prediction head and (b) CGN.}
 	\label{fig:model_CGN_pred}
    \vspace*{-0.6cm}
\end{figure}

To train our RCU, we assume that chunk size is equal to one GOP of size $T$ of 8 frames. For videos with $fps = 25$, the chunk size is 0.32 seconds. We employ the QCIF dataset \cite{Dataset}, comprising videos presented in an uncompressed YUV format. This dataset contains 25 types of videos, all in the QCIF video format, defined by dimensions of $176$ pixels in width $W$ and $140$ pixels in height $H$. We decompose these videos into uniform chunks of 8 consecutive frames, utilizing the FFmpeg tool for this purpose. We further encode all chunked videos with FFmpeg to all 52 video qualities, resulting in a total of 66,768 video chunks. Out of the complete chunk set, we allocate 60,000 chunks for training purposes, while the remaining chunks are reserved for the testing phase. The batch size $B$  considered is 32. In our simulations, we use Adam optimizer with a learning rate of $\eta = 10^{-4}$. The model used for the prediction head and CGN in RCU are shown in Fig.~\ref{fig:model_CGN_pred}. 

We evaluate reconstruction performance with peak signal-to-noise ratio, calculated as PSNR (dB) = $10\log_{10}\left(\frac{\max^2}{MSE}\right)$. 

During the training, the RCU model takes an input pair comprising a video chunk (from the training-set) and its corresponding encoded bitrates and the output of the RCU is compared against the input label comprising the corresponding $QP$ value for the selected training video chunk. We evaluate the performance of the trained RCU model using the test-set. As detailed in Section~\ref{sec:RC}, we address the inherent rounding effect in our RTRC design by incriminating the output $QP$ from the DL RC model for desired accuracy. While the estimated output by DL model yeilds accuracy of $95.7\%$, adjusting the $QP$ value one (two) step higher than the estimated output by the DL model yields an accuracy of $99.12\%$ ($99.87\%$).

\begin{figure}[]
 \centerline{\includegraphics[width=0.78\linewidth]{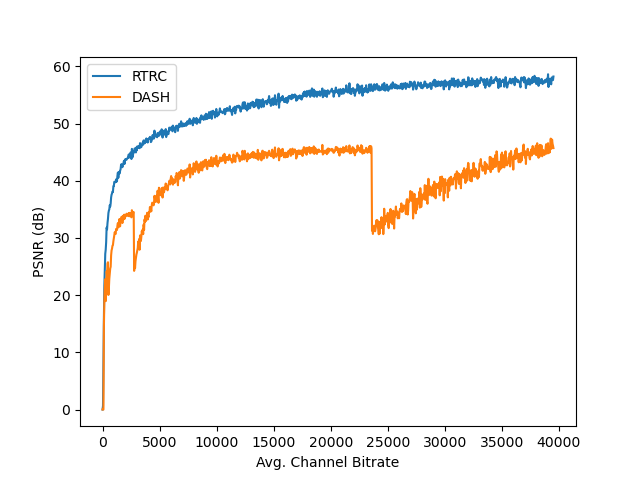}}
 \vspace*{-0.3cm}
  \caption{PSNR vs. average channel bitrate on fading condition.}
  \label{fig:PSNR_vs_ACB}
  \vspace*{-0.5cm}
\end{figure}

Fig.~\ref{fig:PSNR_vs_ACB} provides a comparative analysis of the proposed RTRC framework and DASH algorithm in terms of reconstructed video quality (measured in tems of PSNR) versus the average channel bitrate across scenario of wireless fading channels. The results demonstrate that our novel RTRC method consistently outperforms DASH in all channel circumstances, with particularly pronounced gains in fading scenarios where RTRC showcases  10-20\% higher PSNR values. 

The considerable improvement achieved by our innovative RTRC technology lies within two key attributes. First, the ability of RTRC to dynamically adapt the encoding bitrates encompassing an expansive spectrum of 52 distinct video quality configurations in real-time. In contrast, DASH is confined to a limited assortment of 5 pre-established video quality levels. 

Second, within the RTRC framework, a strategic configuration of $QP$ ensures that allocated bandwidth remains marginally below the predetermined threshold of the channel bitrate. In contrast, DASH adopts an alternative approach, striving to sustain consistent quality until the average channel bitrate surpasses the average bitrate associated with the subsequent available video quality level. 
To elaborate, DASH makes quality determinations grounded in average quality bitrates observed over multiple chunks (i.e., segments lasting 10 seconds). Hence, DASH may suffer from noticeable artifacts due to (i) variation of the bitrates of a segment across its chunks and (ii) variation of the channel bitrate due to fading. 

\begin{figure}[]
 \centerline{\includegraphics[width=0.78\linewidth]{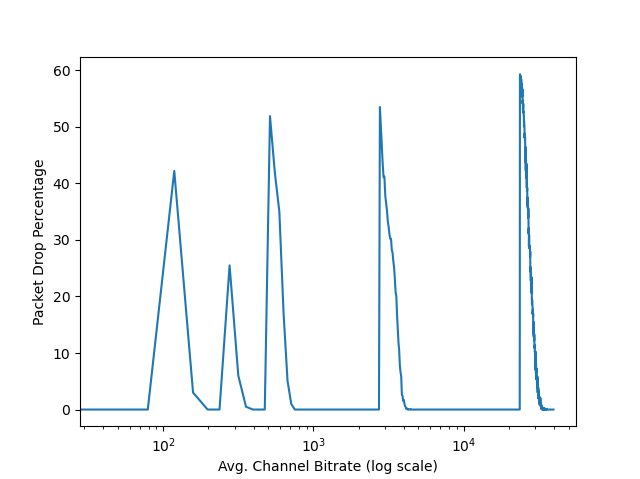}}
 \vspace*{-0.3cm}
  \caption{Packet drop vs. Average channel bitrate on a fading condition.}
  \label{fig:PDP_vs_ACB_fading}
  \vspace*{-0.4cm}
\end{figure}

Fig.~\ref{fig:PDP_vs_ACB_fading} shows the packet drop percentage as a function of average channel bitrate for DASH, indicating the likelihood of video artifacts per chunk. We observe that particularly at the transitions between the next video, the video bitrate is matched with the average channel quality, and hence there is considerable probability that DASH suffers from the packet drop. However, for the channel qualities which fall between the average bitrates of two available DASH resolutions, packet drop probabilities are usually very small and close to zero. These transitional artifacts are visually illustrated in Fig.~\ref{fig:PSNR_vs_ACB}.

We define bandwidth efficiency as the variance between the video bitrate and the available channel bitrate. Meanwhile, we define the packet success rate as the proportion of chunks that do not exhibit artifacts, meaning the encoded video bitrate is lower than the available channel bitrate.

\begin{figure}[]
 \centerline{\includegraphics[width=0.75\linewidth]{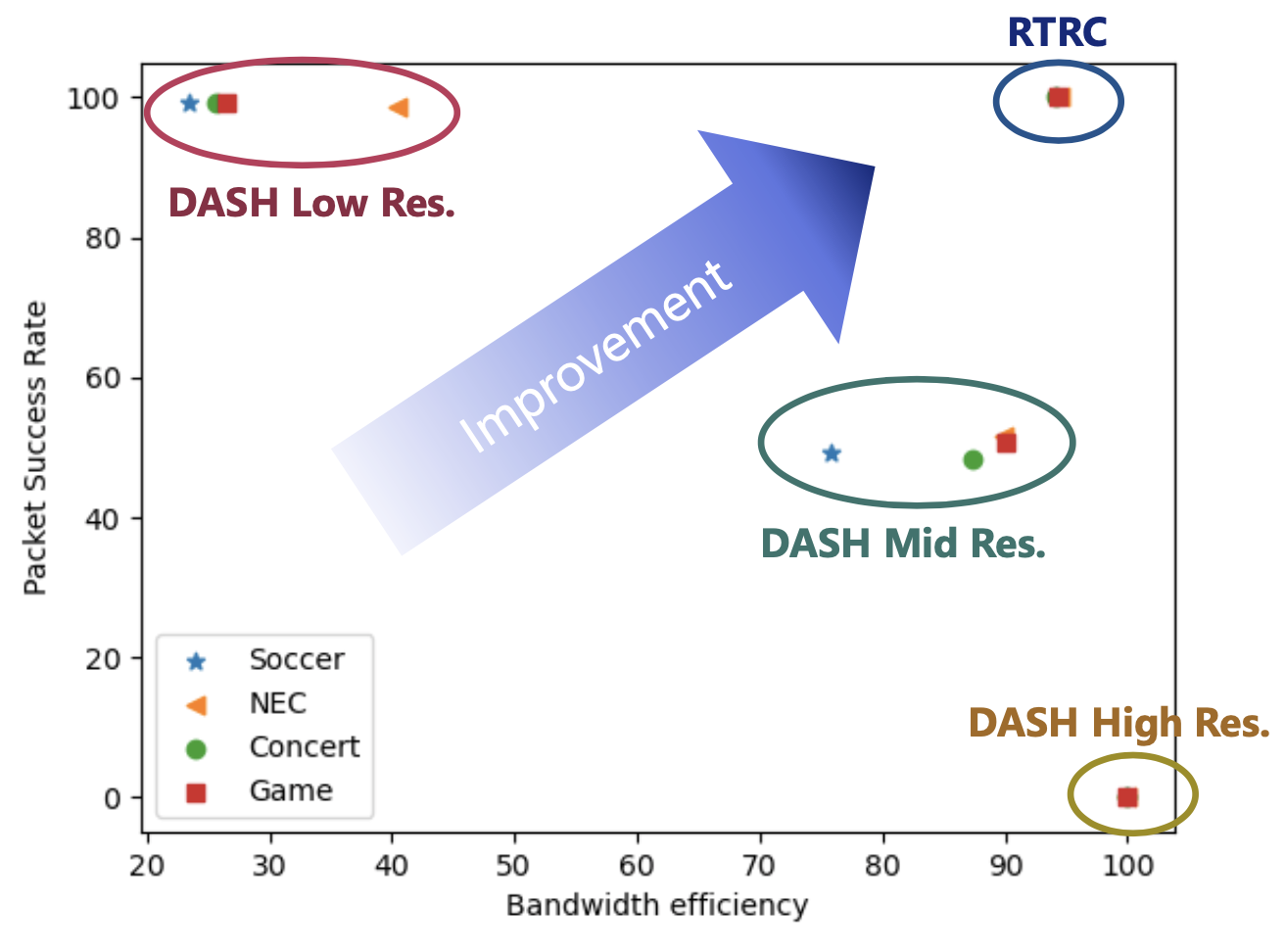}}
 \vspace*{-0.2cm}
  \caption{Packet success rate vs. channel bandwidth efficiency.}
  \label{fig:PSR_vs_ECBE}
  \vspace*{-0.3cm}
\end{figure}

The interplay between packet success rate and channel bandwidth efficiency is shown in Fig.~\ref{fig:PSR_vs_ECBE}. In this experiment, we select four random internet videos and individually encode them at 3 distinct DASH resolutions: high, medium, and low. For all videos, we assume that the available channel bitrate is equal to the average bitrate of DASH's mid-resolution video.

Employing a lower resolution DASH configuration results in nearly 100\% packet success rate because the video bitrate falls below the available channel bitrate. However, this approach underutilizes the available channel bandwidth, achieving only about 25\% bandwidth efficiency. On the other hand, using a higher resolution DASH encoding causes the encoded bitrate to exceed the available channel bitrate, resulting in a 0\% packet success rate but maximizing channel bandwidth utilization.

With mid-resolution DASH encoding, the packet drop rate approaches 40\% due to the closely matched encoded video bitrate and available channel bitrate. However, this approach achieves approximately 80\% bandwidth efficiency.

In contrast, the RTRC outperforms all DASH resolution scenarios where combined effect of packet success rate and channel bitrate utilization is considered. This superiority is due to the strategic selection of an encoded video bitrate just below the available channel bitrate by the RTRC, leading to nearly 99.7\% packet success rate and 95.6\% bandwidth efficiency.

\bibliographystyle{unsrt}
\bibliography{reference}
\end{document}